\documentclass[runningheads]{llncs}
\usepackage[T1]{fontenc}
\usepackage{newtxtext}
\usepackage[varvw]{newtxmath}
\usepackage{graphicx}
\usepackage{cite}
\usepackage{textcomp}
\usepackage{xcolor}
\usepackage{url}
\usepackage[caption=false,font=footnotesize]{subfig}
\usepackage{multirow}
\usepackage{booktabs}
\usepackage{array}
\usepackage{bbding}

\usepackage{hyperref}
\usepackage{color}

\newcommand{\sysname}{\emph{BrainPhone}}

\begin{document}
\title{\sysname{}: A Passive Side-Channel Attack for Speech and Identity Inference Using Consumer BCI Headsets}
\titlerunning{\sysname{}}
\author{Anuradha Mandal\inst{1} \and
Cagri Arisoy\inst{2} \and
Sonjoy Kumar Paul\inst{3} \and
Nitesh Saxena\inst{4}}
\authorrunning{A. Mandal et al.}
\institute{Independent Researcher\\
\email{anuradhamandal.bd@gmail.com} \and
Texas A\&M University, College Station\\
\email{cagri.arisoy@tamu.edu} \and
Texas A\&M University, College Station\\
\email{skpaul@tamu.edu} \and
Texas A\&M University, College Station\\
\email{nsaxena@tamu.edu}}
\maketitle

\begin{abstract}
Consumer brain--computer interfaces (BCIs) that read the electroencephalogram (EEG) are an emerging class of wireless peripherals for gaming, meditation, and productivity. Once paired, they stream continuous raw neural data to applications with no runtime permission prompt and no recording indicator. We present \sysname{}, an end-to-end passive neuro-attack framework through which a malicious, non-privileged application turns that stream into the numeric digits the wearer speaks or hears (PINs, verification codes, account numbers) and the wearer's gender and identity, without access to any audio. The framework works from the raw stream alone and is defined over an arbitrary electrode set: a lightweight speech-event detector locates the moments of speech, a band-limited representation is built for each of them, and sensor-topology-aware models restricted to the electrodes over the brain's speech areas classify them. We evaluate the framework on a commercial 14-channel headset with 20 participants. Models on four frontal electrodes reach 87.85\% accuracy for spoken digits and 86.20\% for listened-to digits (chance 10\%), and the same signals reveal the wearer's gender and identity with 94.97\% and 90.89\% accuracy. The detector, evaluated on utterances it was not trained on, separates 1\,s speech windows from the pauses between utterances with 88.6\% accuracy on the full montage (84.6\% on the same four electrodes) and recovers 77\% of the utterances in unsegmented recordings without any external timing, so the framework does not depend on knowing when a secret is spoken. Headsets with only a few frontal electrodes are not safer than research-grade ones, and the leak sits outside every control that protects the microphone; we derive countermeasures at the operating-system and driver level.

\keywords{Side-channel analysis \and Brain--computer interface (BCI) \and Electroencephalography (EEG) \and Mobile and wearable security \and Speech inference \and Biometric privacy}
\end{abstract}

\section{Introduction}
\label{sec:intro}

Brain--computer interfaces (BCIs) have transformed how people interact with technology. Systems that began as costly clinical tools, restoring movement to people with severe disabilities~\cite{hochberg2012reach}, are now sold as consumer electronics. Electroencephalography (EEG) based BCIs measure the brain's electrical activity non-invasively through a few electrodes on the scalp~\cite{EEGbrainwaves}; long used to diagnose neurological conditions, they are today marketed for gaming and entertainment~\cite{gameEntertain1,gameEntertain2,gameEntertain3}, education~\cite{education1,education2,education3}, and neuromarketing~\cite{neuromarketing1,neuromarketing2,neuromarketing3,neuromarketing4}, bridging cognitive function and everyday digital experiences.

This spread is driven by companies such as NeuroSky~\cite{Neurosky}, Emotiv~\cite{Emotiv}, NextMind~\cite{nextmind}, and Neurable~\cite{Neurable}, which offer low-cost headsets and software development kits through which third-party applications consume the neural stream~\cite{DADEBAYEV20224385}. A growing ecosystem of ``brain apps'' has formed around them, and with it a tension between utility and privacy. On every major mobile platform the microphone and the camera sit behind a runtime permission prompt and a recording indicator. The raw EEG stream of a consumer headset sits behind neither: once the headset is paired, an application obtains continuous brainwave data without any further prompt~\cite{takabi2016brain}. Audits of BCI application stores found no restriction on what an application may do with that stream~\cite{takabi2016brain,bonaci_app_2014}, and a reverse-engineering study showed that every application in the NeuroSky store was open to over-the-air theft of it~\cite{xiaoCanSeeYour2019a}. People wear these headsets for hours during gaming or meditation, and the combination of a seamless experience with minimal ongoing access control leaves them with no sign that their neural activity is being recorded.

The integration of BCIs with AR/VR platforms~\cite{openbci_galea_2025,mullinThisStartupWants2025,azleinClinicalTrialNews2025} and with all-day productivity hardware such as EEG-enabled headphones~\cite{MW75NeuroLT} is rapidly expanding this attack surface. Users of these devices routinely speak sensitive identifiers, PINs, verification codes, meeting passcodes, and payment details, during extended sessions that already carry known side-channel risks from motion and gaze data~\cite{sajidJustStopDoing2025a,pahi_extended_2022,roesner_security_2021}. Privacy scholars have proposed a right to mental privacy~\cite{ienca_towards_2017}, regulators have only recently begun to treat neural data as a category of its own~\cite{yangRegulatingNeuralData2025}, and how far such protection should reach is still debated~\cite{crutchfieldMentalPrivacyCognitive2024}. What that debate lacks is a concrete measure of what an unprivileged application can learn from the stream while its wearer goes about an ordinary day.

Prior work offers two partial answers. Security researchers have shown that a consumer BCI can be turned against its wearer, but only with the adversary's active participation: Martinovic et al.\ flashed images of banks and PIN digits and read the evoked response~\cite{martinovic2012feasibility}, Frank et al.\ hid such probes below the threshold of perception~\cite{frank2017using}, and PEEP decoded keystroke-related activity while the victim typed~\cite{Neupane2017PEEPPE}. None of these applies to a wearer who is simply talking or listening. BCI researchers, on the other hand, have shown that EEG reflects the mental task a person performs~\cite{tan2010brain,del2002local}, the object a person imagines~\cite{esfahani2012classification}, and, with research-grade equipment, the speech a person produces or hears~\cite{krishnaSpeechSynthesisUsing2020,leeFullyEndtoEndListenedSpeech2024a}; but those studies use dense electrode arrays, artifact-controlled protocols, and cooperative subjects, with no adversary in the picture. Between the two lies the premise of this paper: spoken and heard speech engage auditory processing, articulation, and working memory, all of which a consumer headset records, so a passive application should be able to read them from the stream, provided it can find those moments.

We ask: \emph{Can a malicious, non-privileged application use the ``silent'' EEG stream of a consumer BCI as a side channel to infer what its wearer says and hears, and who the wearer is?}

To answer this question we introduce \sysname{}, a neuro-attack framework that turns the raw EEG stream into three predictions: the numeric digits the wearer speaks or hears, the wearer's gender, and the wearer's identity. \sysname{} obtains the stream through any of the access paths above and runs unattended after a one-time setup. A lightweight speech-event detector scans the stream for the seconds that contain speech, which removes the need for the stimulus timestamps that earlier attacks took for granted. Detected windows are decomposed into the canonical EEG bands and classified by sensor-topology-aware models that use only the electrodes over the brain's speech areas, with a majority vote over the samples of each window deciding the digit. The framework is independent of the headset: each stage is defined over an arbitrary electrode set, so the same pipeline can be instantiated on a 14-channel research-style device or on a forehead-only consumer band. We target numeric digits because they are the most directly monetizable spoken secrets, PINs, one-time codes, card and account numbers, and because ten balanced classes make the leakage easy to quantify.

To evaluate the framework, we instantiated it on a commercial 14-channel headset, the Emotiv Epoc X~\cite{epocx}, and recorded 20 participants who spoke and listened to digits for 22,000 trials in total. \sysname{} classifies spoken digits with 87.85\% accuracy and listened-to digits with 86.20\%, against a random-guess baseline of 10\%. The peak accuracy is reached not with the full 14-electrode array but with four frontal electrodes (AF3, F7, F3, FC5), a configuration present in many minimal consumer devices, and a single electrode still yields 72\%. The same four electrodes reveal the wearer's gender with 94.97\% accuracy and the wearer's identity among twenty people with 90.89\%. The speech-event detector, evaluated on utterances it was not trained on, separates 1\,s speech windows from the pauses between utterances with 88.6\% accuracy (84.6\% with the same four electrodes) and recovers 77\% of the utterances in an unsegmented stream at 70\% precision.

This paper moves beyond the general observation that neural data is sensitive to a framework that recovers symbolic information from it, spoken and heard digits read directly from neural signals, and to a measurement of how much a passive attacker gains by doing so. Our contributions are:
\begin{enumerate}
    \item \textbf{A novel passive side channel.} To our knowledge, \sysname{} is the first side channel that infers what a BCI wearer says and hears, and who the wearer is, from the raw EEG of a consumer headset. We define a concrete threat model for neuro-surveillance in everyday settings (Section~\ref{sec:threat-model}). Unlike prior EEG attacks that require visual stimuli~\cite{martinovic2012feasibility,lange2018sidechannel,frank2017using} or observation of keyboard entry~\cite{Neupane2017PEEPPE}, \sysname{} is a fully passive, persistent monitor of the neural stream during normal device use.
    \item \textbf{An end-to-end neuro-attack framework.} \sysname{} is a pipeline that (i) locates speech events in the continuous stream with a compact convolutional detector, (ii) builds a band-limited, per-electrode representation of each event, and (iii) performs supervised inference with sensor-topology-aware classifiers and majority-vote decoding (Section~\ref{sec:Attack-Framework}). Every stage is parameterized by the electrode set, so the framework applies to any headset that exposes its raw stream, and the detector answers the question earlier attacks left open: how an attacker finds the sensitive seconds in hours of EEG without any external timing.
    \item \textbf{An evaluation across electrode configurations.} We instantiate the framework on a 14-channel consumer headset and evaluate it on 20 participants across nested electrode subsets (Sections~\ref{experiment_design} and~\ref{sec:results}). Four left-frontal electrodes outperform the full array on both digit tasks and carry the gender and identity leak, a single electrode retains most of the leak, and the detector survives the same reduction. Devices with only a few forehead electrodes, the cheapest and most common consumer form factor, are therefore not safer than research-grade ones.
    \item \textbf{Quantified real-world risk and countermeasures.} The framework recovers spoken and listened-to digits, gender, and identity from the neural channel alone (Section~\ref{sec:results}), and we show what the per-digit accuracy means for PINs and one-time codes. From where the leak was measured we derive countermeasures at the operating-system, driver, and vendor level (Section~\ref{sec:discussion}).
\end{enumerate}

Section~\ref{sec:background} reviews consumer EEG and prior brainwave attacks, Section~\ref{sec:threat-model} defines the threat model, Section~\ref{sec:Attack-Framework} presents the framework, Section~\ref{experiment_design} describes how we evaluate it, Section~\ref{sec:results} reports the results, and Section~\ref{sec:discussion} discusses limitations and countermeasures.

\section{Background and Related Work}
\label{sec:background}

\subsubsection{EEG and brainwaves.}
EEG records the voltage fluctuations produced by ionic currents in cortical neurons through electrodes placed on the scalp~\cite{schomerNiedermeyersElectroencephalographyBasic2017}. Neural oscillations, or brainwaves, are rhythmic patterns in this activity, characterized by their frequency, amplitude, and phase~\cite{brainwave}, and are linked to perception, motor control, and memory. The signal is conventionally divided into five frequency bands; this work uses delta (0.1--4\,Hz), theta (4--8\,Hz), alpha (8--12\,Hz), beta (12--30\,Hz), and gamma (30--40\,Hz). Each band characterizes a different class of activity: delta dominates deep sleep, theta is associated with imagination and memory, alpha with relaxation, beta with alertness and awareness, and gamma with motor function and mental activity~\cite{EEGbrainwaves}.

\subsubsection{Consumer EEG devices.}
\begin{figure}[!htb]
\centering
\begin{minipage}[b]{0.24\columnwidth}
\centering
\subfloat[]{\includegraphics[width=0.95\linewidth,pagebox=cropbox,trim=24 118 54 42,clip]{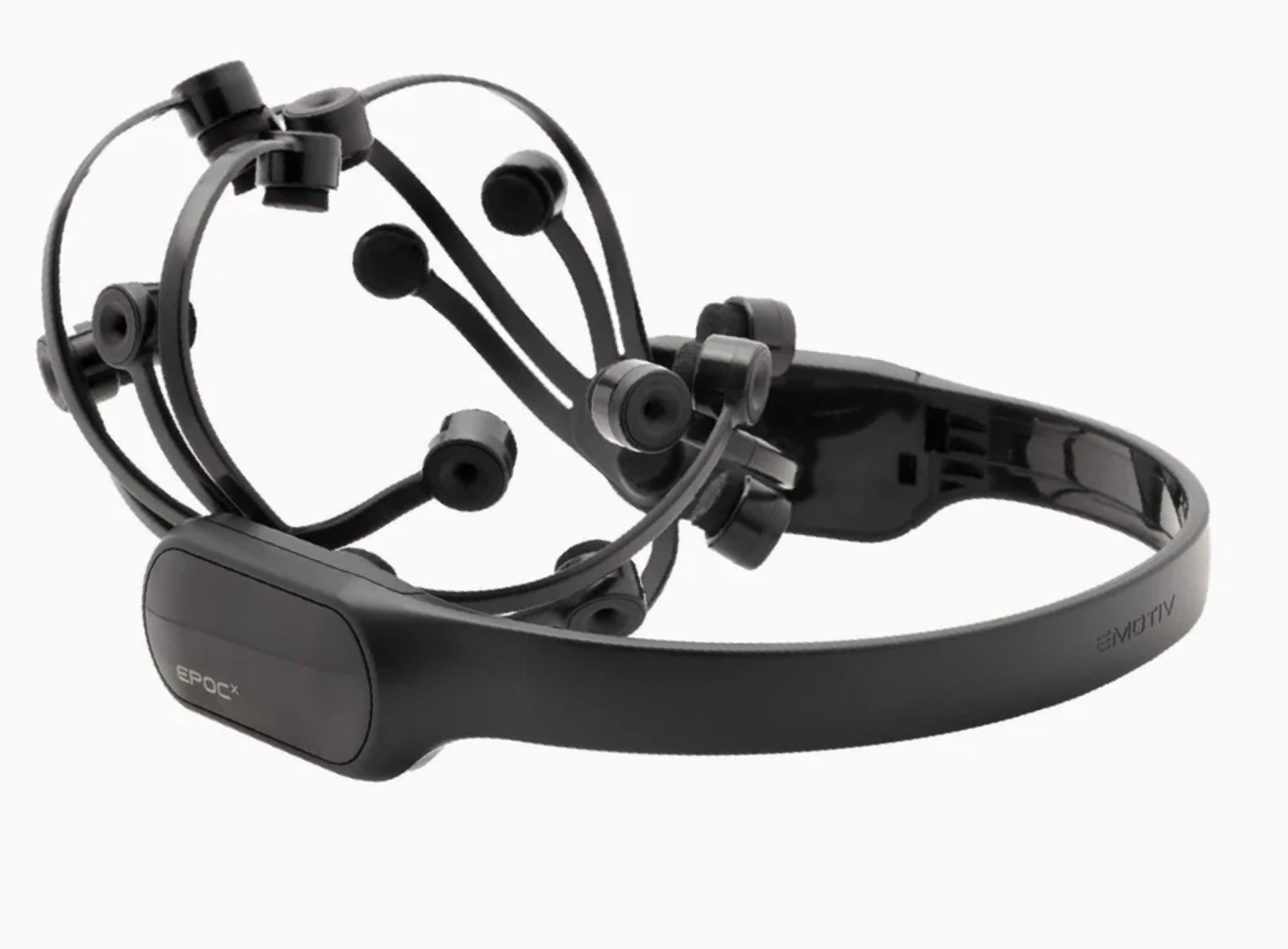}}\\[2pt]
\subfloat[]{\includegraphics[width=0.85\linewidth,pagebox=cropbox]{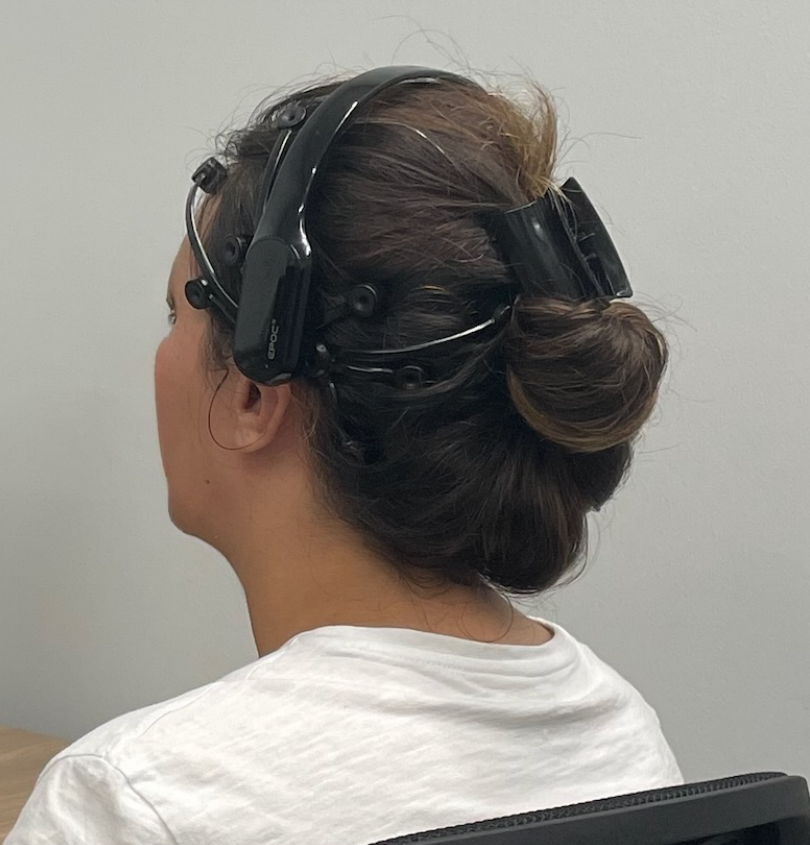}}
\end{minipage}\hspace{0.04\columnwidth}\begin{minipage}[b]{0.39\columnwidth}
\centering
\subfloat[]{\includegraphics[width=\linewidth,pagebox=cropbox,trim=28 26 30 8,clip]{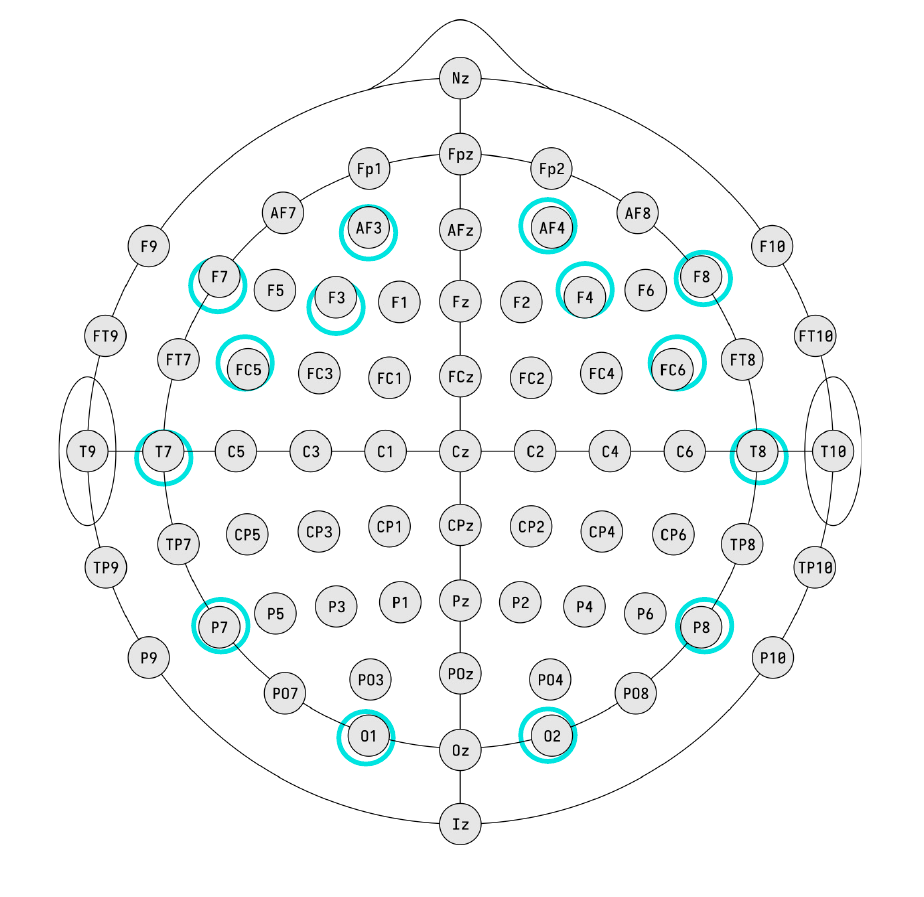}}
\end{minipage}
\caption{The Emotiv Epoc X: (a) the 14-channel wireless headset~\cite{epocx}; (b) a participant wearing it during the experiment; (c) its electrode positions on the international 10--20 layout. The four left-frontal sites of our best models (AF3, F7, F3, FC5) lie over the left inferior frontal cortex.}
\label{fig:eeg}
\end{figure}

Consumer headsets trade electrode count and signal quality for comfort: the Emotiv Epoc X used in this work has 14 saline electrodes at the 10--20 positions AF3, F7, F3, FC5, T7, P7, O1, O2, P8, T8, FC6, F4, F8, and AF4 (Figure~\ref{fig:eeg}), samples at 128\,Hz, and needs no skin preparation~\cite{epocx}. Cheaper devices keep only frontal sites: the Emotiv Insight has five electrodes including AF3 and AF4~\cite{Insight5Channel2025}, and the NeuroSky MindWave has a single forehead electrode~\cite{MindWaveEEGHeadset}. Vendor SDKs expose the stream to third-party code: Emotiv's Cortex API publishes a raw \texttt{eeg} stream to any application whose license carries the \texttt{eeg} scope~\cite{emotivCortexDataSub}, NeuroSky's ThinkGear socket relays raw samples to local clients~\cite{thinkgearsocket}, and browsers expose paired devices through the Web Bluetooth API~\cite{WebBluetoothAPI2025}.

\subsubsection{Privacy attacks through brainwaves.}
Martinovic et al.~\cite{martinovic2012feasibility} were the first to treat a consumer BCI as a side channel: by flashing images of banks, PIN digits, and locations and reading the evoked P300 response, they reduced the entropy of the victim's secrets, reaching 20\% correct PIN-digit guesses. Lange et al.~\cite{lange2018sidechannel} revisited the PIN attack using a 32-electrode research EEG system: after training on each victim's EEG responses, their attack identified the victim's real 4-digit PIN among five random 4-digit codes from 5--40 further observations per code. Frank et al.~\cite{frank2017using} pushed the probe below conscious perception (13.3\,ms). PEEP~\cite{Neupane2017PEEPPE} made the channel passive by decoding keystroke-related activity while the victim typed PINs and passwords, with per-digit true-positive rates of 31--48\% depending on the model. Mandal and Saxena systematized the area~\cite{EEGsok} and demonstrated inference of health conditions~\cite{DiseaseDetector} and news interests~\cite{NewsReader} from single-electrode headsets. Takabi et al.~\cite{takabi2016brain} and Bonaci et al.~\cite{bonaci_app_2014} audited BCI app stores and found no permission layer between applications and raw EEG; Xiao et al.~\cite{xiaoCanSeeYour2019a} showed that the stream can additionally be stolen over the air. Table~\ref{tab:related-comparison} positions \sysname{} against these attacks: it is passive, needs no stimulus, targets the victim's own speech and hearing, and adds identity and gender inference.

\begin{table}[!htb]
\centering
\scriptsize
\caption{EEG privacy attacks. Results are quoted from the respective papers and are not directly comparable; ours come from the protocol of Section~\ref{sec:eval-protocol}.}
\label{tab:related-comparison}
\setlength{\tabcolsep}{3pt}
\renewcommand{\arraystretch}{1.0}
\begin{tabular}{@{}>{\raggedright\arraybackslash}p{2.0cm}>{\raggedright\arraybackslash}p{2.73cm}>{\raggedright\arraybackslash}p{1.68cm}>{\raggedright\arraybackslash}p{1.95cm}>{\raggedright\arraybackslash}p{2.99cm}@{}}
\toprule
\textbf{Attack} & \textbf{Mode} & \textbf{Victim does} & \textbf{Inferred} & \textbf{Reported result} \\
\midrule
Martinovic et al.~\cite{martinovic2012feasibility} & Active (visual probes) & Views probes & PIN digit, bank, location, faces & First guess correct in 20\% of cases (PIN digit) \\ \cmidrule{1-5}
Lange et al.~\cite{lange2018sidechannel} & Active (visual probes) & Views probes & Real PIN among 6 codes; identity & Real PIN identified after 5--40 views per code \\ \cmidrule{1-5}
Frank et al.~\cite{frank2017using} & Active, subliminal ($<$13.3\,ms probes) & Watches a video & Familiar faces & Guessing entropy cut by 20.8\% for 18 of 27 subjects \\ \cmidrule{1-5}
PEEP~\cite{Neupane2017PEEPPE} & Passive & Types & PIN digits, password characters & 31--48\% per-digit true-positive rate \\ \cmidrule{1-5}
Disease Detector~\cite{DiseaseDetector} & Passive & Daily activities & Epilepsy & 82--94\% \\ \cmidrule{1-5}
News Reader~\cite{NewsReader} & Passive & Reads & News category & Up to 99.8\% (3 classes) \\ \cmidrule{1-5}
\sysname{} & Passive & Speaks, listens & Digits, gender, identity & 86--88\% per digit, 93--95\% gender, 83--91\% identity (20-way) \\
\bottomrule
\end{tabular}
\end{table}

\subsubsection{Decoding speech from EEG.}
A large BCI literature decodes \emph{imagined} speech to help users with communication impairments, using CNNs over spectro-spatial inputs~\cite{ko2022spectro}, CNN--LSTM models for imagined characters, digits, and objects~\cite{kumarDeepSpatioTemporalModel2021}, recurrent models for imagined digits~\cite{Mahapatra_2023}, feature-extraction comparisons~\cite{rusnac2022cnn}, metaheuristically tuned CNNs~\cite{rathod2022aquila}, and large public datasets~\cite{vivancos2022mindbigdata}; recent reviews summarize the state of the art~\cite{shah2022role,jin2025systematic}. Work on \emph{overt} and \emph{perceived} speech is closer to our setting: Krishna et al.\ synthesized speech features from EEG recorded during speaking and listening~\cite{krishnaSpeechSynthesisUsing2020}; Lee et al.\ reconstruct listened speech waveforms end-to-end~\cite{leeFullyEndtoEndListenedSpeech2024a}; Xiong et al.\ synthesize intelligible utterances from imagined-speech EEG using overt-speech EEG as a bridge~\cite{xiong2025synthesizing}; NeuroHeed uses EEG to steer speaker extraction toward the attended talker~\cite{pan2024neuroheed}; and Moreira et al.\ release a 64-channel dataset of listened and repeated phonemes, syllables, and words~\cite{moreira2025openaccess}. These results establish that overt and perceived speech are encoded in scalp EEG, but they rely on high-density research equipment, artifact-controlled protocols, and cooperative subjects. \sysname{} asks the complementary security question: how much of that information survives a 14-electrode consumer headset, a 128\,Hz sampling rate, and an adversary who can only obtain labels by deception.

\subsubsection{EEG as a biometric.}
EEG carries stable, person-specific structure: authentication systems reach 97--98\% accuracy, including for users unseen during training~\cite{khalilUnlockingSecurityComprehensive2024,bidgoly2022universal}, and EEG features improve speaker verification under noise~\cite{hanRobustEndtoEndSpeaker2021}. The property that makes EEG a credential makes it a tracker once an unprivileged application holds the stream.

\subsubsection{Speech side channels through other sensors.}
Smartphone motion sensors have been used to recover speech from loudspeaker vibrations~\cite{Gyrophone,Spearphone}, although live human speech alone moves them little~\cite{Speechless}. Those channels are indirect and require the sound source to be co-located with the phone; the EEG channel is direct, follows the wearer, and persists for as long as the headset is worn. Bajwa et al.~\cite{bajwa2019detecting} isolated a single electrode (FC5) from a 14-electrode Emotiv EPOC montage to detect drivers' responses to distracting stimuli, an early hint that sparse frontal montages are exploitable.

\section{Threat Model}
\label{sec:threat-model}

\begin{figure}[!htb]
\centering
\includegraphics[width=0.9\linewidth]{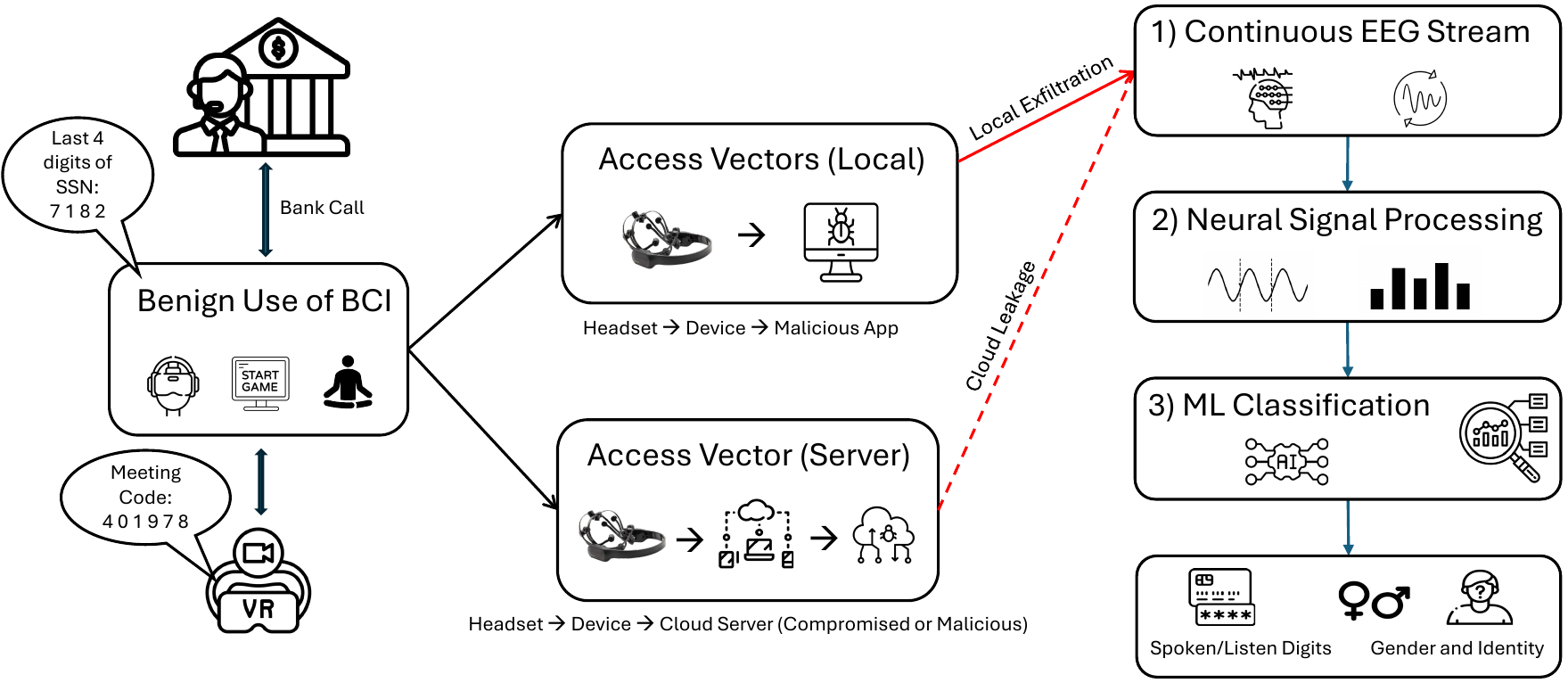}
\caption{High-level overview of the \sysname{} neuro-attack framework. A benign-looking application obtains the raw EEG stream locally or through a cloud backend, runs a speech-event detector on the stream, and classifies detected events into digits, gender, and identity.}
\label{fig:sp-attack-engine}
\end{figure}

We define a practical, software-based threat model in which an adversary with only application-level access eavesdrops on the wearer's acoustic environment by analyzing raw EEG.

\subsubsection{Setting.}
The victim wears a consumer EEG headset (e.g., Emotiv Epoc X or Insight~\cite{epocx,Insight5Channel2025}, NeuroSky MindWave~\cite{MindWaveEEGHeadset}, Neurable MW75 Neuro~\cite{MW75NeuroLT}, OpenBCI Galea~\cite{openbci_galea_2025}) paired over Bluetooth with a phone, PC, or AR/VR headset that runs third-party companion applications. Three deployments ground the threat. \emph{Workplace EEG headphones} such as the MW75 Neuro are designed for all-day wear during which users take calls and read out verification codes; workers already experience physiological monitoring at work as a deep privacy intrusion~\cite{roemmich_emotion_2023}. \emph{BCI gaming} sessions last hours and run alongside voice chat in which players read account codes and make purchases. \emph{AR/VR with integrated EEG} combines neural sensing with teleconferencing, where users speak meeting passcodes in an environment whose own security and privacy risks are already an active research area~\cite{roesner_security_2021,pahi_extended_2022,sajidJustStopDoing2025a}.

\subsubsection{Adversary and goals.}
The adversary controls software that receives the raw EEG stream but has no privileges beyond those of an ordinary application: no OS exploit, no Bluetooth protocol break, and, crucially, no microphone permission. Its goals are (G1)~\emph{speech inference}, recovering numeric sequences the victim speaks or hears (PINs, one-time codes, card, account, and social-security numbers), and (G2)~\emph{profiling}, learning the wearer's gender and re-identifying the wearer across sessions and applications for tracking, de-anonymization, or targeted social engineering.

\subsubsection{Access vectors.}
\emph{Local compromise.} A trojanized ``brain game,'' wellness tracker, or productivity utility requests the standard Bluetooth permission, pairs with the headset, and records raw EEG whenever the headset is active. Vendor SDKs make this a supported use: Emotiv's Cortex API exposes a raw \texttt{eeg} stream to any application whose developer license includes that scope~\cite{emotivCortexDataSub}, NeuroSky's ThinkGear socket relays raw samples to every local client~\cite{thinkgearsocket}, and app-store audits found no per-application restriction on the stream~\cite{takabi2016brain,bonaci_app_2014}. A web page can achieve the same with zero installation through the Web Bluetooth API after a single pairing consent~\cite{WebBluetoothAPI2025,CommunicatingBluetoothDevices}. \emph{Cloud leakage.} Many companion applications upload raw EEG to backends for ``AI-powered'' analytics; a compromised or dishonest operator can run \sysname{} offline over archived streams. \emph{Over-the-air theft.} Xiao et al.\ demonstrated that home-use EEG streams can be captured by a nearby attacker without touching the victim's device~\cite{xiaoCanSeeYour2019a}. \sysname{} is agnostic to which vector delivers the stream.

\subsubsection{Why not just use the microphone?}
An attacker with microphone access would not need EEG. The point of the neural channel is that it sidesteps the controls that mobile platforms have built around audio: runtime permission prompts, persistent recording indicators, and, in managed deployments, corporate policies that block microphone access in third-party applications. A Bluetooth peripheral's data stream has none of them: pairing is consented to once, and nothing shows that samples are being read afterwards. We therefore treat EEG as a \emph{permission-bypassing} channel rather than a replacement for audio. It is also a cheap channel. The Epoc X emits 14 channels of at most 16-bit samples at 128\,Hz, about 3.5\,kB/s or 13\,MB per hour, a tenth of uncompressed 16\,kHz audio, and the four electrodes our best models use amount to under 4\,MB per hour; the detector is a network of about 5,000 parameters and the classifiers are random forests, so both run on the phone and only detected windows, or the digits decoded from them, need leave the device. Finally, the channel yields something the microphone does not: a physiological identifier of the wearer (G2) that persists even when the wearer is silent.

\subsubsection{Passive, persistent surveillance.}
\sysname{} runs as a background monitor for as long as the headset is worn. This sets it apart from earlier EEG attacks, which needed the adversary to act during the attack (show a probe~\cite{martinovic2012feasibility}, or wait for typing~\cite{Neupane2017PEEPPE}). Most of a long recording is uninteresting, so the attacker has to find the seconds that matter. \sysname{} does so in two stages: a lightweight \emph{speech-event detector} flags windows whose EEG looks like speech (Section~\ref{sec:event-detection}), and OS-level context (an active VoIP call, the dialer or a banking application in the foreground, a meeting-PIN prompt in AR) narrows the search to high-value moments. Detected windows then go to the digit, gender, and identity classifiers. Figure~\ref{fig:sp-attack-engine} summarizes the pipeline. The only interaction with the wearer is a one-time setup, in which the application requests a short prompted recording of the kind BCI and voice-assistant onboarding flows already include~\cite{MentalCommandsEmotiv,emotivBCITraining,NeurableFAQs,PersonalizedHeySiri}; gender and identity inference (G2) need no such step, only EEG collected under a known account.

\subsubsection{Out of scope.}
We do not attack the Bluetooth link, the headset firmware, or the OS.

\section{The \sysname{} Neuro-Attack Framework}
\label{sec:Attack-Framework}

\sysname{} is an end-to-end framework that turns a raw EEG stream into three predictions: the digit being spoken or heard, the wearer's gender, and the wearer's identity. It has four stages: (1)~signal preprocessing and band decomposition, (2)~speech-event detection in the continuous stream, (3)~feature representation of detected windows, and (4)~sensor-topology-aware classification with majority-vote decoding. Three requirements from the threat model shape the design. The framework must be \emph{passive}: it consumes the stream as the application receives it and, after a one-time setup, never prompts the wearer. It must be \emph{self-timed}: nothing tells the attacker when a secret is spoken, so the framework has to find the moments itself. And it must be \emph{headset-agnostic}: consumer devices differ in electrode count and placement, so every stage is defined over an arbitrary electrode set $E$, from the full montage of a research-style headset down to a single forehead electrode, and the same pipeline runs unchanged on each. We describe the stages in general terms and give the parameters of our instantiation on the 14-channel Emotiv Epoc X, which Section~\ref{sec:results} evaluates over nested subsets of $E$.

\subsection{Preprocessing and Band Decomposition}
\label{sub-sec:signal-processing}
The input is the multichannel stream at the headset's native sampling rate, after whatever filtering the device applies on board; for the Epoc X this is 14 channels at 128\,Hz behind a fifth-order sinc filter (effective bandwidth 0.16--43\,Hz) and 50/60\,Hz notch filters~\cite{epocx}. We decompose each channel into the five canonical bands with zero-phase band-pass filters (delta 0.1--4, theta 4--8, alpha 8--12, beta 12--30, gamma 30--40\,Hz), producing a $5|E|$-dimensional band-limited time series at the original sampling rate, 70 dimensions for the full Epoc X montage. Each participant-session is z-scored per dimension so that amplitude differences between people and between sessions (anatomy, electrode impedance, headset fit), the subject-specific variability that EEG decoders must contend with~\cite{jinInterpretableCrossSubjectEEGBased2020}, do not dominate the features.

\subsection{Finding Speech in the Stream}
\label{sub-sec:localization}
An attacker who records for hours does not know when a digit was spoken. Two observations make the problem tractable: speech leaves a mark in consumer EEG that is visible by eye, and a small network can learn that mark from a few minutes of recording. We illustrate the first with one participant's recording (Figures~\ref{fig:time-domain}--\ref{fig:one-spectra} in the supplementary material) and then describe the detector.

\emph{What speech looks like in the stream.} Figure~\ref{fig:time-domain} contrasts a rest segment with five spoken digits: each utterance is a burst of high-amplitude activity, and the pauses between utterances are visible in the raw trace. The event-related spectral perturbation at FC5 (Figure~\ref{fig:time-freq}) shows that the bursts are broadband, spanning delta through gamma, rather than confined to one band. Band-power topographies concentrate beta and gamma power around FC5, FC6, and F4 (Figure~\ref{fig:one-spectra}). Speech events are thus separable from background EEG, and the frontal sites carry most of the difference, which motivates both the detector below and the electrode subsets evaluated in Section~\ref{sec:results}.

\subsubsection{The speech-event detector.}
\label{sec:event-detection}
\begin{figure}[!htb]
\centering
\includegraphics[width=0.7\columnwidth]{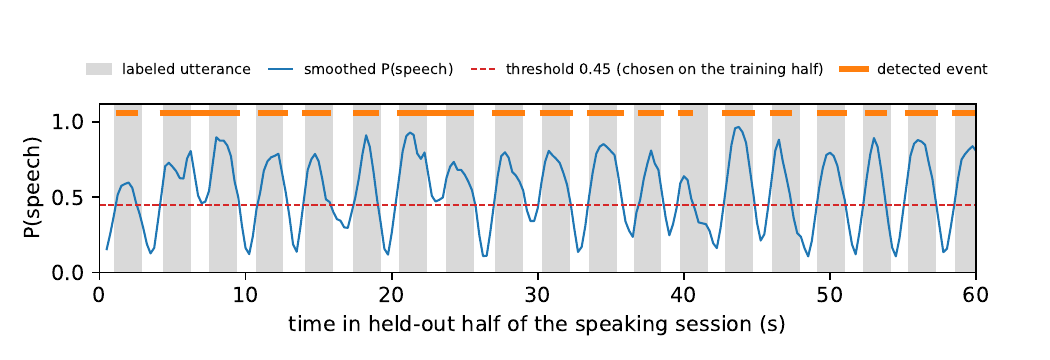}
\caption{Detector output over 60\,s of the held-out second half of one participant's recording (P4); the detector was trained on the first half. The smoothed per-window probability follows the utterances (grey); runs above the threshold chosen on the training half become detected events (orange).}
\label{fig:detector-trace}
\end{figure}

The detector decides, for each window of the stream, whether it contains speech-related activity. It is EEGNet~\cite{lawhernEEGNetCompactConvolutional2018}, a compact convolutional network built for small EEG datasets, trained as a binary classifier on the band-passed (0.5--45\,Hz) signal. Positive examples are fixed-length windows that start at a labeled utterance onset in the training recording; negatives are windows of the same length taken from the pauses between utterances. Windows are 1.0\,s (128 samples), the longest length that still fits into the $\approx$1.1\,s pauses, and they are never zero-padded, since a padded tail would trivially mark the class. At attack time the detector slides over the unsegmented stream with 75\% overlap, its per-window scores are smoothed, and runs above a threshold become events. A few minutes of recording supply everything the detector needs: they contain both speech and the pauses between utterances, so the detector is trained on them and the smoothing window and threshold are chosen on them as well. Figure~\ref{fig:detector-trace} shows the result on a stretch of recording the detector has not seen: the probability rises with every utterance and falls in the pause after it, and thresholding the trace recovers the utterances one by one.

\subsection{Feature Representation}
\label{sub-sec:features}
The representation has to satisfy three constraints that follow from the threat model. It must be computable on the victim's phone from the raw stream, it must work with the short windows the detector hands over, and it must be learnable from a short recording. The band-limited series of Section~\ref{sub-sec:signal-processing} meet all three. From each detected (or, in our controlled evaluation, labeled) window we use a fixed-length segment, 119 samples ($\approx$0.93\,s) for speaking and 61 samples ($\approx$0.48\,s) for listening, the length covered by every utterance and every stimulus in our evaluation data, so that all windows contribute equally. Every sample inside the segment then becomes one feature vector: the instantaneous amplitude of each of the five bands at each electrode, 70 values for the full montage, 20 for the four left-frontal electrodes, and five for AF3 alone. Vectors are z-scored per participant and recording, which removes the amplitude differences caused by electrode impedance and headset fit and leaves the spatio-spectral pattern of the moment: which bands are active, and where.

A window thus yields 119 (or 61) vectors that share a label, and over our evaluation data this gives 238,000 vectors for the 2,000 spoken digits and 1.2\,million for the 20,000 heard ones. We chose one vector per sample over per-window summary statistics (band powers, spectral entropies, or independent components) for four reasons. It needs no windowing heuristics beyond the detector's output; it costs one filter bank per electrode, well within a phone's budget; it lets the classifier learn from every sample of a short recording rather than from one summary per utterance; and it lets the decoder vote over many observations of the same utterance (Section~\ref{sub-sec:classification}), which absorbs the errors of individual samples. The same vectors serve all three targets; the gender and identity models attach a different label to them.
\subsection{Sensor-Topology-Aware Classification}
\label{sub-sec:classification}

\begin{table}[!htb]
\centering
\scriptsize
\caption{Classifier variants. Each is a Random Forest over the band-limited features of the listed electrodes.}
\label{tab:mlmodels}
\setlength{\tabcolsep}{4pt}
\begin{tabular}{@{}llll@{}}
\toprule
\textbf{Model} & \textbf{Region} & \textbf{Electrodes} & \textbf{Target} \\
\midrule
\textit{global-\{spoken,listen\}} & All lobes & all 14 & Digit \\
\textit{left-\{spoken,listen\}} & Left hemisphere & AF3, F7, F3, FC5, T7, P7, O1 & Digit \\
\textit{front-\{spoken,listen\}} & Frontal lobe & AF3, F7, F3, FC5, FC6, F4, F8, AF4 & Digit \\
\textit{left-front-\{spoken,listen\}} & Left frontal & AF3, F7, F3, FC5 & Digit \\
\textit{onesensor-spoken} & Single site & AF3 & Digit \\
\textit{gender-\{spoken,listen\}} & Left frontal & AF3, F7, F3, FC5 & Gender \\
\textit{user-\{spoken,listen\}} & Left frontal & AF3, F7, F3, FC5 & Identity \\
\bottomrule
\end{tabular}
\end{table}

A central hypothesis of this work is that electrode placement matters more than electrode count, because the information the attacker wants is produced by specific cortical regions and additional channels add noise. The classifier stage therefore takes the electrode set $E$ as a parameter, and we instantiate it as a family of Random-Forest classifiers on nested electrode subsets (Table~\ref{tab:mlmodels}); each subset also approximates the electrode layout of a cheaper headset with fewer channels. The \textit{global} models use all 14 electrodes. The \textit{left} models use the seven left-hemisphere sites, following the left lateralization of language~\cite{ries2016choosing,friederici_brain_2011}. The \textit{front} models use the eight frontal sites, which host most executive and speech-planning function~\cite{collins2012reasoning,alexander1989frontal}. Their intersection, the four-electrode \textit{left-front} models (AF3, F7, F3, FC5), sits over the left inferior frontal gyrus, including Broca's area, which is central to speech production and articulation~\cite{ucsfSpeechLanguage2025,friederici_brain_2011}. A \textit{onesensor} model (AF3 only) approximates single-electrode consumer devices. The same four-electrode subset is reused for gender and identity.

We chose Random Forests for their robustness to the non-Gaussian, heavy-tailed distribution of band-limited EEG samples and for their negligible inference cost on a mobile device. Each forest has 100 trees with a maximum depth of 20 for speaking and 25 for listening, selected under 10-fold cross-validation. \emph{Decoding.} Each sample in a window receives a class label; the window's label is the majority vote over its samples. Since a PIN or code is a sequence of such windows, the sequence is recovered by concatenating window-level votes.

\section{Evaluating the Framework}
\label{experiment_design}
To evaluate \sysname{} we instantiated it on one consumer headset and collected speaking and listening EEG from 20 participants. This section describes the participants, the two tasks, the handling of their data, and how each stage of the framework is scored.

\subsection{Participants and Apparatus}
We recruited 20 participants through internal university e-mail, a sample in line with prior EEG security studies (28 in~\cite{martinovic2012feasibility}, 27 in~\cite{frank2017using}, 12 in~\cite{Neupane2017PEEPPE}): 8 female and 12 male by self-reported gender, 13 aged 25--34, six 35--44, and one 45 or older (Appendix~\ref{app:demographics}). All self-reported normal hearing. Participants wore an Emotiv Epoc X headset (Figure~\ref{fig:eeg}) fitted without skin preparation, exactly as a consumer would, while seated in front of a computer. Consent and data handling are described in Section~\ref{sec:ethics}.

\subsection{Tasks}
\begin{enumerate}
    \item \textbf{Speaking.} Digits 0--9 were shown one at a time on the screen and the participant read each aloud. Each of ten blocks presented all ten digits in random order, for 100 utterances per participant (2,000 in total). An utterance lasted about 1.9\,s from prompt to the next prompt.
    \item \textbf{Listening.} The participant listened to pre-recorded digits spoken by ten speakers (5 female, 5 male; Appendix~\ref{app:demographics}). Each speaker--digit pair was played five times in random order with a 500\,ms gap, for 500 stimuli per session; every participant completed two listening sessions on the same day, for 1,000 stimuli per participant (20,000 in total).
\end{enumerate}
Prompted presentation of digits is the standard protocol in EEG digit- and speech-recognition studies~\cite{Mahapatra_2023,imagenetofbrain2018mindbigdata,vivancos2022mindbigdata,kumariConvolutionalNeuralNetworkBased2021}. Only EEG and prompt timestamps were recorded; \emph{no audio was captured at any point}, so all results rest on the neural channel alone.

\subsection{Ethical Considerations}
\label{sec:ethics}
The study was approved by our university's Institutional Review Board. Participants gave written informed consent after an explanation of the headset, the data collected (raw EEG, prompt timestamps, and self-reported demographic attributes), the purpose of the study, and the privacy risks under investigation, and could withdraw at any time. Only the digits 0--9 were spoken or heard; no real PINs, passwords, or account numbers were elicited. Recordings are indexed by opaque identifiers, stored on encrypted institutional servers, and will not be released in raw form; the analysis pipeline and de-identified aggregate features will be. Because this work demonstrates an offensive capability, we weighed disclosure against harm: the signal-processing primitives are already in the literature~\cite{martinovic2012feasibility,frank2017using,Neupane2017PEEPPE}, consumer BCIs already expose raw EEG to third-party applications~\cite{takabi2016brain,emotivCortexDataSub}, and the countermeasures we propose require vendor and platform action that is unlikely without a demonstrated attack. We judge publication to be net-beneficial.

\subsection{Evaluation Protocol}
\label{sec:eval-protocol}
\subsubsection{Digit, gender, and identity classifiers.}
For each task and electrode set we train one classifier on the pooled, per-participant z-scored feature vectors of all 20 participants (238,000 for speaking, 1.2\,M for listening); there is no separate model per participant. This is the model the attacker of Section~\ref{sec:threat-model} holds: one model trained on the setup recordings of all its users, so that each wearer's own recording is part of the training set, as the identity task requires in any case. We report two evaluations, 10-fold cross-validation and a single 80\%/20\% train--test split, both over the pooled feature vectors, which are the unit of classification (Section~\ref{sub-sec:features}); unless stated otherwise, numbers in the text are from the held-out split. Metrics are computed per feature vector: true-positive rate (TPR), false-positive rate (FPR), macro precision, recall, F-measure, and accuracy; the trial-level decision is the majority vote of Section~\ref{sub-sec:classification}. Chance is 10\% for digits, 50\% for gender, and 5\% for identity.

\subsubsection{Speech-event detector.}
We evaluate the detector in two settings of increasing difficulty, each for the full montage and for the two electrode subsets of the digit classifiers (the four left-frontal electrodes; AF3 alone). (i)~\emph{Pre-segmented}: a trial-disjoint 80\%/20\% split and, separately, 5-fold cross-validation over all windows. The classes are 1\,s windows at utterance onsets and 1\,s windows centered in the pauses between utterances of the same block, so both classes come from the same behavioral state, seconds apart. We report EEGNet and, as a check that the separability does not hinge on one architecture, a gradient-boosted-tree classifier over per-channel band statistics. (ii)~\emph{Unsegmented}: one detector per participant is trained on the first half of the session and slid over the second half with 75\% overlap and without any timestamps; detections are merged into events and matched to true utterances by temporal overlap (IoU $\geq$ 0.3), which yields precision, recall, onset error, and false alarms per minute. The smoothing window, threshold, and merge gap are chosen per participant on the training half, as an attacker would have to choose them in advance.

\section{Results}
\label{sec:results}

We evaluate the framework stage by stage. We first report the digit, gender, and identity classifiers (Sections~\ref{sec:res-spoken}--\ref{sec:res-biometric}), then the speech-event detector on pre-segmented and unsegmented data (Section~\ref{sec:res-detection}), and close with a consolidated summary (Section~\ref{sec:results-generalization}).

\subsection{Spoken Digits}
\label{sec:res-spoken}
\begin{table}[!htb]
\centering
\scriptsize
\caption{Spoken-digit classification (10 classes, chance 10\%; accuracy in \%). Metrics over held-out feature vectors, pooled over 20 participants.}
\label{tab:digit-recog-spoken}
\setlength{\tabcolsep}{2pt}
\begin{tabular}{@{}l cccccc cccccc@{}}
\toprule
& \multicolumn{6}{c}{\textbf{10-fold cross-validation}} & \multicolumn{6}{c}{\textbf{80\%/20\% held-out split}} \\
\cmidrule(lr){2-7}\cmidrule(lr){8-13}
\textbf{Model} & TPR & FPR & Prec & Rec & FM & Acc. & TPR & FPR & Prec & Rec & FM & Acc. \\
\midrule
\textit{global-spoken} (14) & 0.754 & 0.027 & 0.754 & 0.754 & 0.754 & 75.43 & 0.743 & 0.029 & 0.743 & 0.743 & 0.743 & 74.26 \\
\textit{left-spoken} (7) & 0.860 & 0.016 & 0.860 & 0.860 & 0.860 & 86.01 & 0.850 & 0.017 & 0.850 & 0.850 & 0.850 & 84.95 \\
\textit{front-spoken} (8) & 0.839 & 0.018 & 0.839 & 0.839 & 0.839 & 83.92 & 0.820 & 0.020 & 0.820 & 0.820 & 0.820 & 81.97 \\
\textit{left-front-spoken} (4) & 0.889 & 0.012 & 0.889 & 0.889 & 0.889 & 88.94 & 0.879 & 0.014 & 0.879 & 0.879 & 0.879 & \textbf{87.85} \\
\textit{onesensor-spoken} (1) & 0.737 & 0.029 & 0.737 & 0.737 & 0.737 & 73.69 & 0.722 & 0.031 & 0.722 & 0.722 & 0.722 & 72.24 \\
\bottomrule
\end{tabular}
\end{table}

Table~\ref{tab:digit-recog-spoken} shows the spoken-digit results. The \textit{global-spoken} model over all 14 electrodes reaches 74.26\%, already $7.4\times$ chance, yet it is the \emph{weakest} multi-electrode configuration. Restricting to the left hemisphere raises accuracy to 84.95\%, restricting to the frontal lobe to 81.97\%, and their four-electrode intersection, \textit{left-front-spoken} (AF3, F7, F3, FC5), reaches the best result of \textbf{87.85\%}, an $8.8\times$ improvement over chance. The per-class metrics move together: on the held-out split the four-electrode model has a true-positive rate of 0.879 and a false-positive rate of 0.014, so a feature vector of one digit is mistaken for any particular other digit about once in seventy. Cross-validation and held-out numbers agree within one to two points for every model, with the same ordering under both.

The ordering is consistent with the neuroanatomy of speech. Production is left-lateralized, and the four best electrodes lie over the left inferior frontal cortex that hosts Broca's area~\cite{ries2016choosing,ucsfSpeechLanguage2025,friederici_brain_2011}. Adding electrodes outside that region does not add information; it adds noise. The global model includes ten sites that contribute activity unrelated to the digit, occipital electrodes that reflect visual processing of the on-screen prompt and right-hemisphere sites with little involvement in speech production, and the forest's discriminative power drops accordingly. The security consequence is direct: a headset with only forehead electrodes, the montage of low-cost focus and productivity devices~\cite{MindWaveEEGHeadset,MW75NeuroLT}, is not a weaker attack surface than a research-grade array. Even a single electrode (\textit{onesensor-spoken}, AF3) retains 72.24\%, about two points below the full 14-channel model, so one well-placed frontal electrode captures most of the leak.

What a per-digit accuracy means for a secret depends on the secret's length. If digit errors were independent, 87.85\% per digit would recover a 4-digit PIN in full with probability $0.8785^4 = 60\%$ and a 6-digit one-time code with 46\%, and the true 6-digit code would lie within one substitution of the decoded string, among 55 candidates rather than $10^6$, with probability 84\%. Errors within a session are unlikely to be independent, so these figures are indicative, but they show that even partial recovery collapses the search space of a long number by orders of magnitude, and that an attacker who observes a code twice, read aloud and then repeated to an agent, gains a second draw.

\subsection{Listened-to Digits}
\label{sec:res-listen}
\begin{table}[!htb]
\centering
\scriptsize
\caption{Listened-digit classification (10 classes, chance 10\%; accuracy in \%). Metrics over held-out feature vectors, pooled over 20 participants and two sessions.}
\label{tab:digit-recog-listen}
\setlength{\tabcolsep}{2pt}
\begin{tabular}{@{}l cccccc cccccc@{}}
\toprule
& \multicolumn{6}{c}{\textbf{10-fold cross-validation}} & \multicolumn{6}{c}{\textbf{80\%/20\% held-out split}} \\
\cmidrule(lr){2-7}\cmidrule(lr){8-13}
\textbf{Model} & TPR & FPR & Prec & Rec & FM & Acc. & TPR & FPR & Prec & Rec & FM & Acc. \\
\midrule
\textit{global-listen} (14) & 0.730 & 0.030 & 0.730 & 0.730 & 0.712 & 72.96 & 0.717 & 0.031 & 0.717 & 0.717 & 0.717 & 71.73 \\
\textit{left-listen} (7) & 0.847 & 0.017 & 0.847 & 0.847 & 0.847 & 84.68 & 0.828 & 0.019 & 0.828 & 0.828 & 0.828 & 82.83 \\
\textit{front-listen} (8) & 0.870 & 0.014 & 0.870 & 0.870 & 0.870 & 86.97 & 0.851 & 0.017 & 0.851 & 0.851 & 0.851 & 85.08 \\
\textit{left-front-listen} (4) & 0.878 & 0.014 & 0.878 & 0.878 & 0.878 & 87.85 & 0.862 & 0.015 & 0.862 & 0.862 & 0.862 & \textbf{86.20} \\
\bottomrule
\end{tabular}
\end{table}

The listening task (Table~\ref{tab:digit-recog-listen}) follows the same pattern: 71.73\% with all electrodes, 82.83\% with the left hemisphere, 85.08\% with the frontal lobe, and \textbf{86.20\%} with the four left-frontal electrodes ($8.6\times$ chance). Two properties of the listening data make this result stronger than the number alone suggests. Each digit was spoken by ten different voices, five female and five male, and the label is the digit rather than the speaker, so the classifier has to recognize the heard digit across the acoustic variability of ten talkers. And the listening windows are half the length of the speaking windows (0.48\,s against 0.93\,s), so every decision rests on fewer samples. Listening still trails speaking by only 1.65 points.

That gap has a natural explanation: speaking adds motor-planning and auditory-feedback activity (and, as discussed in Section~\ref{sec:discussion-emg}, articulatory EMG) to the perceptual response, whereas listening yields the perceptual response alone. The frontal model slightly outperforms the left-hemisphere model for listening, the reverse of the speaking task, consistent with the more bilateral organization of auditory and semantic processing in comprehension~\cite{friederici_brain_2011}; the four-electrode intersection nonetheless remains the most informative region for both tasks. The practical implication is that the attacker need not wait for the victim to speak: a code read \emph{to} the victim over a call, a verification code an agent reads back, or a meeting passcode announced in a voice chat leaks nearly as well as one the victim reads out, and a victim guards a conversation's listening side least.

\subsection{Gender and Identity}
\label{sec:res-biometric}
\begin{table}[!htb]
\centering
\scriptsize
\caption{Gender (2 classes, chance 50\%) and identity (20 classes, chance 5\%) classification from the four left-frontal electrodes (accuracy in \%). Metrics over held-out feature vectors.}
\label{tab:gender-user-recog}
\setlength{\tabcolsep}{2pt}
\begin{tabular}{@{}l cccccc cccccc@{}}
\toprule
& \multicolumn{6}{c}{\textbf{10-fold cross-validation}} & \multicolumn{6}{c}{\textbf{80\%/20\% held-out split}} \\
\cmidrule(lr){2-7}\cmidrule(lr){8-13}
\textbf{Model} & TPR & FPR & Prec & Rec & FM & Acc. & TPR & FPR & Prec & Rec & FM & Acc. \\
\midrule
\textit{gender-spoken} & 0.950 & 0.050 & 0.950 & 0.950 & 0.950 & 94.96 & 0.950 & 0.050 & 0.950 & 0.950 & 0.950 & \textbf{94.97} \\
\textit{gender-listen} & 0.942 & 0.058 & 0.942 & 0.942 & 0.942 & 94.16 & 0.933 & 0.067 & 0.933 & 0.933 & 0.933 & 93.27 \\
\textit{user-spoken} & 0.913 & 0.005 & 0.912 & 0.913 & 0.912 & 91.25 & 0.909 & 0.005 & 0.909 & 0.909 & 0.909 & \textbf{90.89} \\
\textit{user-listen} & 0.837 & 0.009 & 0.837 & 0.837 & 0.837 & 83.68 & 0.834 & 0.009 & 0.834 & 0.834 & 0.834 & 83.42 \\
\bottomrule
\end{tabular}
\end{table}

Table~\ref{tab:gender-user-recog} reports the profiling goals (G2) with the same four electrodes. For \textit{gender-spoken}, we balanced the classes by using 16 participants (8 female, 8 male); the listening data were already balanced through the 5/5 speaker set. The wearer's gender is inferred at \textbf{94.97\%} from speaking EEG, $1.9\times$ chance for a binary decision, with cross-validation and held-out results within 0.01 points of each other. The \textit{speaker's} gender is inferred at 93.27\% from the \emph{listener's} EEG: the neural response to a male versus a female voice is discriminable, so the channel leaks an attribute of the conversation's other party, not only of the wearer.

Identity is recovered at \textbf{90.89\%} among 20 wearers from speaking EEG ($18\times$ chance) and 83.42\% from listening EEG ($16.7\times$), with a per-class false-positive rate of 0.005 in the twenty-way problem; that is, a feature vector from one wearer is attributed to any particular other wearer once in two hundred. These figures agree with the EEG-biometrics literature, where person-specific structure supports 97--98\% authentication~\cite{khalilUnlockingSecurityComprehensive2024,bidgoly2022universal}, and they turn that structure into a liability. For the attacker, identity is the most convenient target of all: it needs only EEG collected under a known account and no detector, and it works from listening EEG, i.e., while the wearer is silent. The same stream that leaks \emph{what} was said therefore reveals \emph{who} said or heard it.
\subsection{Speech-Event Detection}
\label{sec:res-detection}
\begin{table}[!htb]
\centering
\scriptsize
\caption{Speech-event detection by electrode set (chance 50\%). Pre-segmented: trial-disjoint 80\%/20\% split, 5-fold cross-validation in parentheses; 1\,s speech windows vs.\ the pauses between utterances. Unsegmented: one detector per participant slid over the second half of the session, operating point chosen on the first half, detections matched to utterances.}
\label{tab:event-detection}
\setlength{\tabcolsep}{2pt}
\renewcommand{\arraystretch}{1.0}
\begin{tabular}{@{}llccc@{}}
\toprule
\textbf{Setting} & \textbf{Metric} & \textbf{All 14} & \textbf{AF3, F7, F3, FC5} & \textbf{AF3} \\
\midrule
Pre-segmented & EEGNet acc.\ (\%) & \textbf{88.57} (86.0 $\pm$ 1.2) & 84.55 (82.1 $\pm$ 1.1) & 82.08 (80.8 $\pm$ 1.0) \\
 & Boosted-tree acc.\ (\%) & 81.56 (82.9 $\pm$ 1.5) & 81.69 (81.0 $\pm$ 1.3) & 78.70 (78.1 $\pm$ 1.0) \\
\midrule
Unsegmented & Precision / recall (\%) & \textbf{69.5 / 77.2} & 65.4 / 67.3 & 71.6 / 75.5 \\
 & Mean onset error (s) & 0.37 & 0.38 & 0.34 \\
 & False alarms / min & 5.6 & 5.8 & 4.9 \\
\bottomrule
\end{tabular}
\end{table}

{Table~\ref{tab:event-detection} shows how well the detector finds speech without external timing. The negative class is the pause \emph{between} utterances of the same block, a second away from the positives in the same behavioral state, not rest EEG from before the task. On pre-segmented data, EEGNet separates speech from pause with 88.57\% accuracy (5-fold cross-validation 86.0 $\pm$ 1.2\%), and a gradient-boosted-tree classifier over band statistics reaches 81.56\%. On unsegmented EEG, one detector per participant with its operating point chosen on the training half (Figure~\ref{fig:detector-trace}) recovers 77.2\% of the utterances at 69.5\% precision, with a mean onset error of 0.37\,s and 5.6 false alarms per minute over 61 minutes of held-out recording; with the four left-frontal electrodes it still recovers 67.3\% of the utterances at 65.4\% precision, and with AF3 alone 75.5\% at 71.6\%. A PIN read aloud is a burst of utterances, so an attacker who catches three of every four onsets has candidate windows for the digit classifier from the raw stream alone.

\subsection{Summary of Findings}
\label{sec:results-generalization}

\begin{table}[!htb]
\centering
\footnotesize
\caption{Summary of \sysname{} accuracy with the four left-frontal electrodes (AF3, F7, F3, FC5), held-out split.}
\label{tab:results-summary-table}
\begin{tabular}{@{}lccc@{}}
\toprule
\textbf{Inference task} & \textbf{Acc.\ (\%)} & \textbf{Chance (\%)} & \textbf{Gain} \\
\midrule
Spoken digit (10-class) & 87.85 & 10.0 & $8.8\times$ \\
Listened digit (10-class) & 86.20 & 10.0 & $8.6\times$ \\
Wearer gender, speaking (2-class) & 94.97 & 50.0 & $1.9\times$ \\
Speaker gender, listening (2-class) & 93.27 & 50.0 & $1.9\times$ \\
Wearer identity, speaking (20-class) & 90.89 & 5.0 & $18.2\times$ \\
Wearer identity, listening (20-class) & 83.42 & 5.0 & $16.7\times$ \\
Speech vs.\ pause (2-class) & 84.55 & 50.0 & $1.7\times$ \\
\bottomrule
\end{tabular}
\end{table}

Table~\ref{tab:results-summary-table} brings the results together, and four properties of the framework stand out. The four left-frontal electrodes carry the leak for every task, so electrode \emph{placement}, not count, determines exposure. The framework works whether the victim is the speaker or the listener. One stream leaks three kinds of information at once, content, a demographic attribute, and identity; the identity classifier in particular is, by construction, a tracker of any user the application has seen once. And the detector removes the need for external timing, so the attacker can work from the raw stream alone. These findings say where a defense must sit: on the raw stream, at the application boundary (Section~\ref{sec:countermeasures}).

\section{Discussion}
\label{sec:discussion}

\subsection{Scope and Limitations}
\label{sec:limitations}

\emph{Laboratory conditions.} Participants sat still and performed one task at a time. Walking, typing, chewing, and concurrent cognitive load all add EEG variance that we have not measured; the detector's false-alarm rate in particular will rise with movement, for which context gating (Section~\ref{sec:threat-model}) is the attacker's natural response. All recordings of a participant were made on one day, so the results do not measure robustness to a headset re-seated days later.

\emph{Single device.} We evaluated one headset. The four-electrode result suggests transfer to devices covering the left-frontal region, including the five-channel Emotiv Insight (AF3, AF4) and single-electrode NeuroSky devices (\textit{onesensor-spoken}, 72\%), but sampling rate, electrode material, and on-board filtering differ across vendors.

\emph{Vocabulary and non-numeric speech.} We decode the ten digits. Numeric secrets are the highest-value target (PINs are 4--6 digits, one-time codes 6, card numbers 16), but a 26-letter or open-vocabulary attack would face a larger label space and, judging from the imagined-speech literature surveyed in~\cite{jin2025systematic}, lower accuracy. Two-digit numbers (``twenty'') and continuous speech would additionally stress segmentation. Most of what a wearer says is not a digit at all, and the framework has no reject class: the detector flags speech, not digits, and the classifier assigns a digit to every window it receives. In ordinary conversation this yields spurious digits, which the attacker must filter with the context signals of Section~\ref{sec:threat-model} (an active call, a banking application in the foreground) and with the structure of the target, since a code is a burst of four to six short utterances. Our evaluation elicited only digits, so the framework's false-digit rate on unconstrained speech is unmeasured.

\subsection{Neural Signal or Muscle Artifact?}
\label{sec:discussion-emg}
Articulation produces facial and jaw EMG that is orders of magnitude larger than cortical EEG and is picked up preferentially by frontal electrodes; the history of overt-speech EEG is one of ``neural'' effects that dissolved once EMG was controlled for~\cite{devos2010removal}. We therefore do not claim that the speaking-task results are purely neural; the \textit{left-front-spoken} model almost certainly benefits from a myogenic component. For an \emph{attacker} that is immaterial, and it is why PEEP likewise counts muscle activity as part of its channel~\cite{Neupane2017PEEPPE}. The listening task, however, involves no articulation, and there the same four electrodes reach 86.20\% digit accuracy and 93.27\% speaker-gender accuracy. The listening results are our evidence that the channel carries cognitive information beyond EMG; a definitive attribution would require simultaneous facial EMG recording, which we leave to future work.
\subsection{Countermeasures}
\label{sec:countermeasures}
The results point to where a defense must sit: between the headset and unprivileged code, before raw samples are exposed. We propose four measures at that boundary and say what our measurements imply for each; evaluating them empirically is future work.
\begin{enumerate}
    \item \textbf{Semantic permissions instead of raw access.} Operating systems and BCI drivers should expose \emph{derived} signals (``focus level,'' a discrete mental command) through a permission model like the microphone's, and reserve raw EEG for a separately granted, indicator-backed permission~\cite{takabi2016brain,chizeck2014brain,maiseli_braincomputer_2023}. Users do not connect a sensor to what can be inferred from it and judge a request by how the derived information is described~\cite{schneegass_understanding_2019}, so the permission has to be phrased at the level of what is inferred. Feature extraction on the headset or in a privileged driver removes the raw stream from the application boundary; computing a few band-power features locally costs little, and secure multiparty computation over users' EEG~\cite{agarwal2019protecting} fits the same boundary. This does not require cloud-oriented products to become on-device ones: the companion application and its backend keep receiving the derived signals they consume, at a fraction of the raw stream's bandwidth; only the raw samples stop crossing the application boundary.
    \item \textbf{Context-aware minimization.} The OS already knows when a call is active, when the microphone is open, and when a password field has focus. During such events it should suspend or heavily down-sample the EEG stream to third-party applications, exactly the moments our threat model targets. Because the leak is confined to the second around each utterance and the detector relies on broadband 0.5--45\,Hz structure, suspension for the few seconds of a code exchange removes both the windows the classifier needs and the cue the detector uses, and is imperceptible to benign applications that track slow states.
    \item \textbf{Driver-level noise.} Defenses against motion-sensor and keystroke-timing side channels add transparent noise at the driver~\cite{shrestha2016slogger,schwarzetal2018keydrown}. A BCI driver can likewise add randomized noise to the raw stream, tuned to break the classification of individual samples while preserving the coarse band-power trends that legitimate applications consume. Our results indicate where the noise is most valuable: the left-frontal channels and the beta/gamma bands. Evaluating the utility--privacy trade-off of such noise is an open problem that our released pipeline makes testable.
    \item \textbf{Enrollment hygiene.} Platforms can require that any prompt asking the user to \emph{speak} or \emph{type} specific content while EEG is streaming goes through a system dialog that names the data collected, as speaker-enrollment flows do today. This does not stop gender or identity inference and therefore complements, rather than replaces, the first three measures.
\end{enumerate}

\section{Conclusion}
\label{sec:conclusion}
This paper presented \sysname{}, an end-to-end passive neuro-attack framework for an emerging class of consumer brain--computer interface devices. The framework lets an adversary with a malicious application and no special privileges turn the raw EEG stream of a consumer headset into sensitive spoken and listened-to numeric speech, as well as the wearer's gender and identity, without any audio. Our evaluation of the framework on a 14-channel consumer headset with 20 participants demonstrates the attack: up to 87.85\% accuracy for spoken digits and 86.20\% for listened-to digits, 94.97\% for gender, and 90.89\% for identity among twenty wearers. Every stage is defined over an arbitrary electrode set, and the framework works with four frontal electrodes and even a single one, its speech-event detector finds the relevant seconds in unsegmented recordings without the oracle timing that earlier EEG attacks assumed, and the listening results show that the channel carries cognitive information beyond muscle artifacts. The attack runs entirely outside the controls that protect the microphone, and we call on the mobile and systems security community to close this gap with permissions for derived signals rather than raw samples, suspension of the stream while a call is active or a secret is being entered, and driver-level noise, before EEG-enabled headphones and headsets become everyday wearables.

\begin{credits}
\subsubsection{\ackname} This research was funded by National Science Foundation grant numbers CNS-2154507 and OAC-2139358.

\subsubsection{\discintname} The authors have no competing interests to declare that are relevant to the content of this article.
\end{credits}

\bibliographystyle{splncs04}
\bibliography{bci-eeg}

\appendix
\section{Speech in the Raw EEG Stream}
\label{app:speech-figures}
\noindent\emph{Supplementary material.} Figures~\ref{fig:time-domain}--\ref{fig:one-spectra} illustrate Section~\ref{sub-sec:localization} with one participant's recording.

\begin{figure}[!htb]
\centering
\includegraphics[width=0.82\columnwidth]{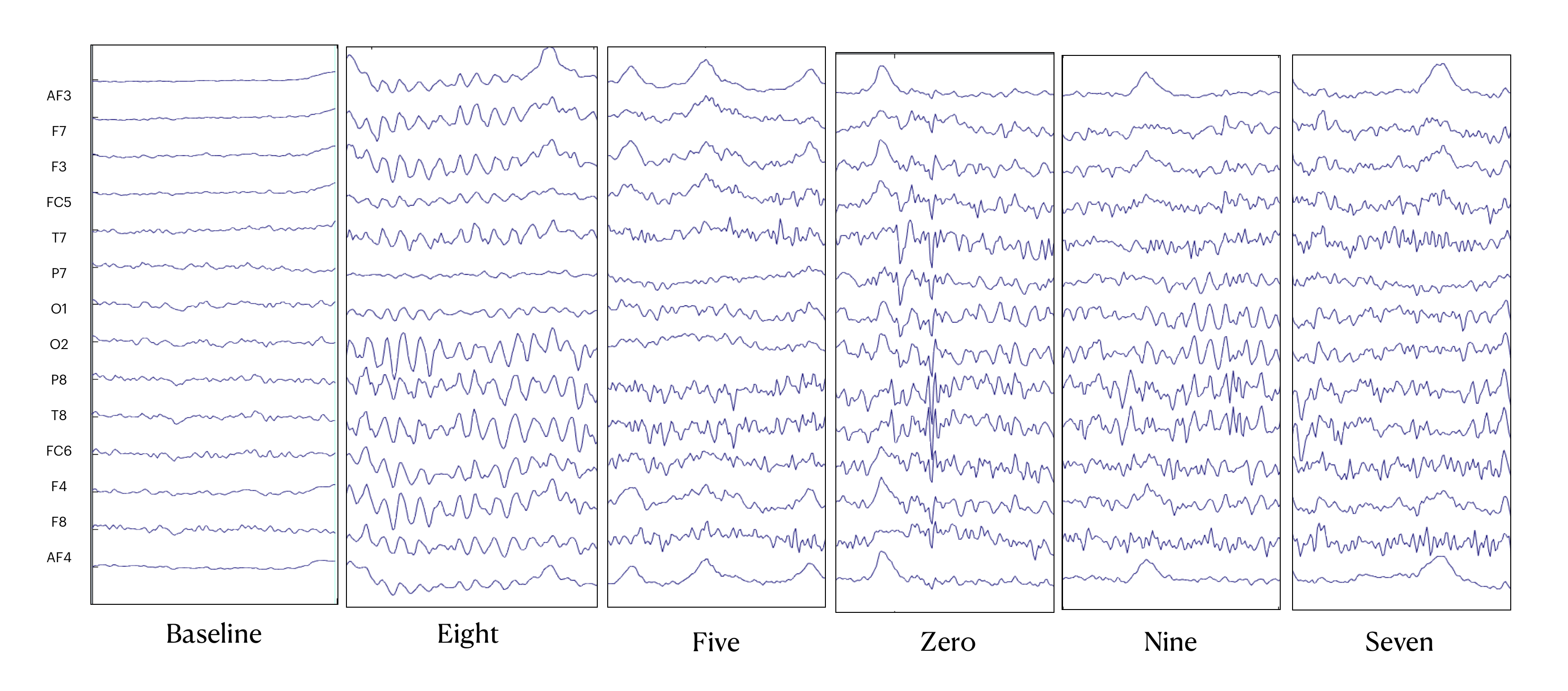}
\caption{Raw 14-channel EEG during rest (left) and five spoken digits. Speech produces broadband, rhythmic amplitude increases across channels relative to the baseline.}
\label{fig:time-domain}
\end{figure}

\begin{figure}[!htb]
\centering
\subfloat[Baseline]{\includegraphics[width=0.44\columnwidth]{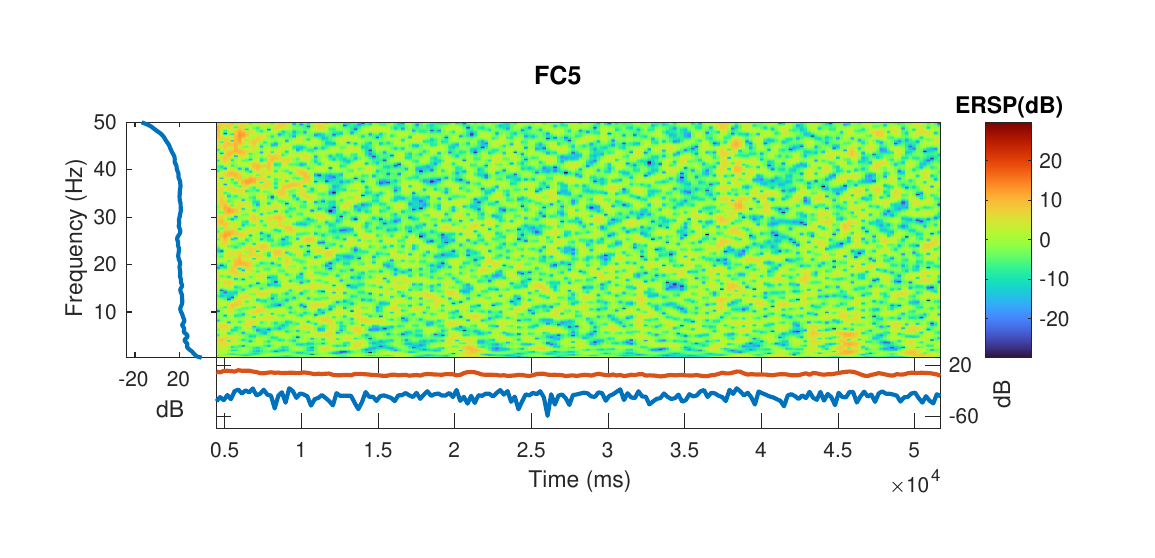}}\hfill
\subfloat[Speaking]{\includegraphics[width=0.44\columnwidth]{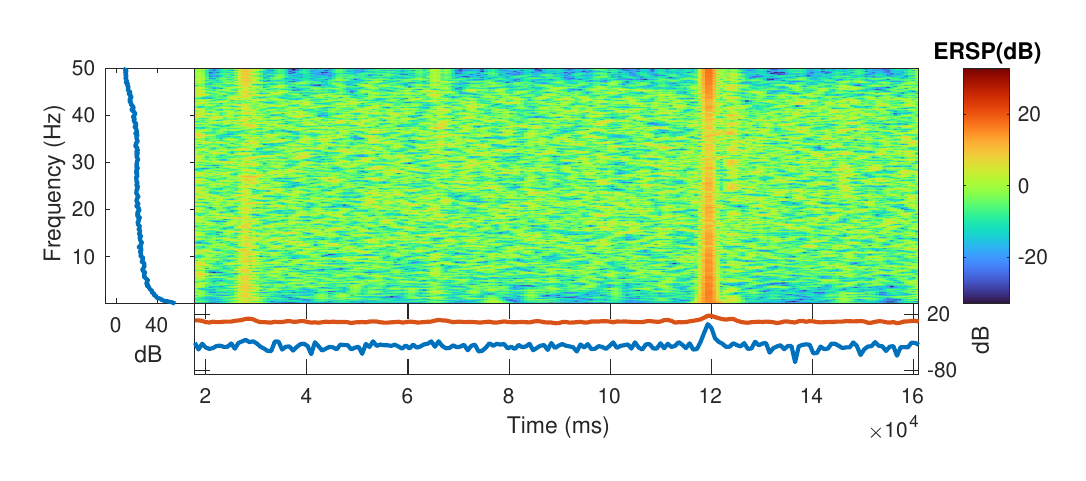}}
\caption{Event-related spectral perturbation at FC5 for baseline (a) and speaking (b). Speech appears as broadband power bursts spanning 0.1--50\,Hz (marginal spectra at left).}
\label{fig:time-freq}
\end{figure}

\begin{figure}[!htb]
\centering
\includegraphics[width=0.52\columnwidth,trim=25 22 12 30,clip]{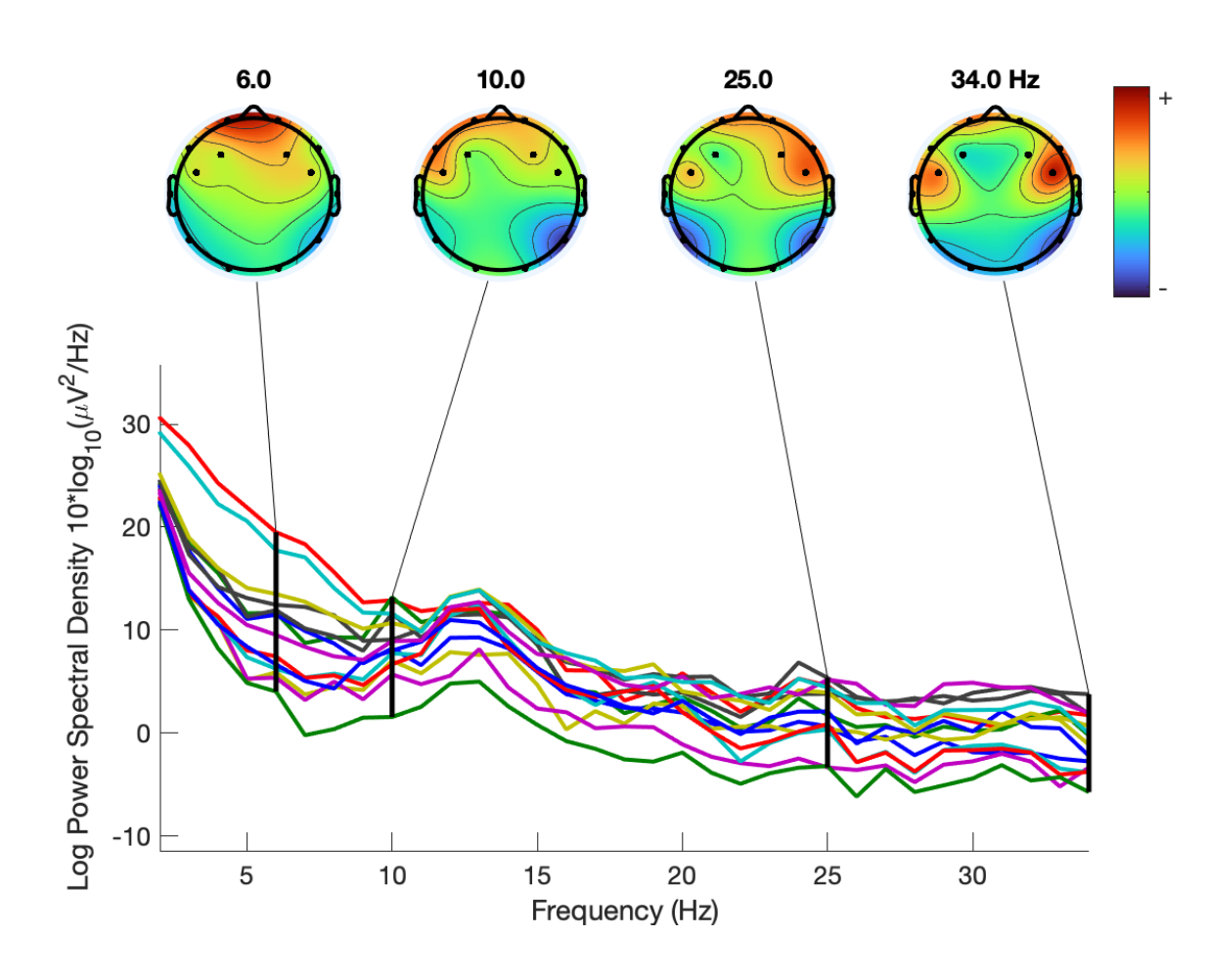}
\caption{Channel spectra (traces) and scalp distribution of power at 6, 10, 25, and 34\,Hz (theta, alpha, beta, gamma) for one spoken-digit session. Beta and gamma power concentrate frontally.}
\label{fig:one-spectra}
\end{figure}

\clearpage
\section{Participant and Speaker Demographics}
\label{app:demographics}
\noindent\emph{Supplementary material.} This appendix supports Section~\ref{experiment_design}.

\begin{table}[!htb]
\centering
\footnotesize
\caption{Participants' age range and gender.}
\begin{tabular}{@{}ccc@{\hspace{2em}}ccc@{}}
\toprule
\textbf{Participant} & \textbf{Age} & \textbf{Gender} & \textbf{Participant} & \textbf{Age} & \textbf{Gender} \\
\midrule
1 & 25--34 & Male & 11 & 25--34 & Male \\
2 & 25--34 & Male & 12 & 35--44 & Male \\
3 & 25--34 & Male & 13 & 35--44 & Male \\
4 & 45+ & Female & 14 & 25--34 & Male \\
5 & 25--34 & Female & 15 & 35--44 & Male \\
6 & 25--34 & Male & 16 & 35--44 & Female \\
7 & 25--34 & Male & 17 & 35--44 & Male \\
8 & 25--34 & Female & 18 & 35--44 & Male \\
9 & 25--34 & Female & 19 & 25--34 & Female \\
10 & 25--34 & Female & 20 & 25--34 & Female \\
\bottomrule
\end{tabular}
\end{table}

\begin{table}[!htb]
\centering
\footnotesize
\caption{Gender of the ten recorded speakers used in the listening task.}
\begin{tabular}{@{}cc@{\hspace{2em}}cc@{}}
\toprule
\textbf{Speaker} & \textbf{Gender} & \textbf{Speaker} & \textbf{Gender} \\
\midrule
1 & Female & 6 & Male \\
2 & Male & 7 & Female \\
3 & Female & 8 & Male \\
4 & Male & 9 & Female \\
5 & Female & 10 & Male \\
\bottomrule
\end{tabular}
\end{table}

\end{document}